\documentclass[aps,prl,twocolumn]{revtex4-1}
\usepackage{bm}%
\expandafter\ifx\csname package@font\endcsname\relax\else
 \expandafter\expandafter
 \expandafter\usepackage
 \expandafter\expandafter
 \expandafter{\csname package@font\endcsname}%
\fi

\usepackage{array}
\usepackage{amsmath,amssymb}
\usepackage{MnSymbol}
 \usepackage{tikz-feynman} 

\usepackage{graphicx}
\usepackage{subfigure}
\usepackage{mathrsfs}
\usepackage{bm}
\usepackage{mathtools}
\usepackage{epstopdf}
\usepackage{hyperref}

\hypersetup{%
   pdfpagemode=None, 
   pdfstartpage=1,
   pdfmenubar=true,
   pdftoolbar=true,
   colorlinks = true,
   linkcolor=blue,
   citecolor=blue,
   urlcolor=blue,
   bookmarksopen=false
 }

\usepackage[normalem]{ulem}   
\definecolor{darkgreen}{rgb}{0,0.60,.2}

\begin{document}
\title{Dropping an impurity into a Chern insulator: a polaron view on topological matter}
\author{A.\ Camacho-Guardian$^1$, N.\ Goldman$^2$, P.\ Massignan$^3$, and G.\ M.\ Bruun$^1$}
\affiliation{$^1$Department of Physics and Astronomy, Aarhus University, Ny Munkegade, DK-8000 Aarhus C, Denmark}
\affiliation{$^2$Center for Nonlinear Phenomena and Complex Systems,
Universit\'e Libre de Bruxelles, CP 231, Campus Plaine, 1050 Brussels, Belgium}
\affiliation{$^3$Departament de F\'isica, Universitat Polit\`ecnica de Catalunya, Campus Nord B4-B5, E-08034 Barcelona, Spain}
\begin{abstract} We investigate the properties of an impurity particle interacting with a Fermi gas in a Chern-insulating state. The interaction leads to the formation of an exotic polaron, which consists of a coherent superposition of the topologically-trivial impurity and the surrounding topological cloud.  We characterize this intriguing topologically-composite object by calculating its transverse (Hall) conductivity, using diagrammatic as well as variational methods. The ``polaronic Hall conductivity", {i.e., the transverse drag exerted by the dressing cloud on the impurity,
is shown to exhibit a sharp jump from zero to a finite value whenever the surrounding cloud enters a topologically non-trivial  state}. In this way, the polaron partially inherits the topological properties of the Chern insulator through genuine interaction effects.
This is also analysed at the microscopic level of wave functions, by identifying a ``composite Berry curvature" for the polaron, which closely mimics the Berry curvature of the Chern insulator's band structure. Finally, we discuss 
how this interplay between topology and many-body correlations can be studied in cold-atom experiments, using available technologies. 
 \end{abstract}
\date{\today}

\maketitle

{\it Introduction.-} The exploration of topological states of matter constitutes one of the most active fields in condensed-matter physics~\cite{Hasan2010,Qi2011,Bernevig2013}.  In parallel to the identification of novel topological properties in single-particle band structures~\cite{bradlyn2017topological,schindler2018higher,schindler2018higher2}, intense efforts are dedicated to the rich interplay of topological bands and inter-particle interactions~\cite{Sorensen2005,Raghu2008,Sun2009,Varney2010,Gurarie2011,Varney2011,cooper2013reaching,Hohenadler2013,Wang2014,Chiu2016,Strinati2017,Repellin2017,Dong2018,Rachel2018}.
In addition to condensed-matter systems, 
topological band structures have also been studied in the context of ultracold gases~\cite{Goldman2016,Cooper2018,Zhang2018}. 
These atomic systems are particularly well suited to investigate the role of interactions in topological 
phases~\cite{Rachel2010,Cocks2012,Kumar2016,Vanhala2016,Tai2017,Salerno2018,Jian2018}, 
since the interaction strength between neutral atoms can be easily tuned experimentally~\cite{Chin2010}. 
 This  feature of cold-atom systems has led to various fundamental 
discoveries~\cite{bloch2008many,bloch2012quantum,gross2017quantum}; in particular, 
ultracold gases have deepened our understanding of how mobile impurities behave within ultracold Fermi~\cite{Schirotzek2009,Kohstall2012,Koschorreck2012,Scazza2017} or Bose~\cite{Jorgensen2016,Hu2016} gases. 

 Here, we show that impurity physics  provides a promising framework to explore interacting topological systems in a realistic and controlled setting. Specifically, we consider an impurity   moving in a honeycomb lattice and interacting with a gas of ``majority" particles forming a Chern insulator. The interaction leads to the formation of a polaron consisting of a (topologically-trivial) impurity dressed by a cloud of majority particles forming a topological phase. Using both a diagrammatic and a variational approach, we calculate the transverse  (Hall) conductivity 
 of this intriguing composite 
 object, and  show that it  partially reflects the Hall-type properties of the majority particles. Physically, this is due to the drag exerted by the dressing 
 cloud on the impurity, and it is thus a genuine interaction effect. At the microscopic level, we
  identify a  ``composite Berry curvature" for the polaron and show that it closely mimics the Berry curvature of the  underlying Chern insulator's band structure.  Our developments 
  are reminiscent of a recently proposed interferometric scheme, which involves mobile impurities bound to quasiparticles in fractional quantum Hall states~\cite{Grusdt2016}.

{\it System.-} Consider a mobile impurity,  denoted as a  $\downarrow$ particle,  immersed in a  gas of fermionic majority (spin $\uparrow)$ particles. 
Both the impurity and the majority particles reside in a honeycomb lattice with nearest neighbour hopping.  In addition, 
the majority particles experience next-nearest neighbour hopping, which breaks time-reversal symmetry, and  a broken inversion symmetry 
given by an energy offset between neighbouring sites, see Fig.~\ref{LatticeFig}(a). 
\begin{figure}[h!]
\begin{center}
\includegraphics[width=3.0in,height=1.5in]{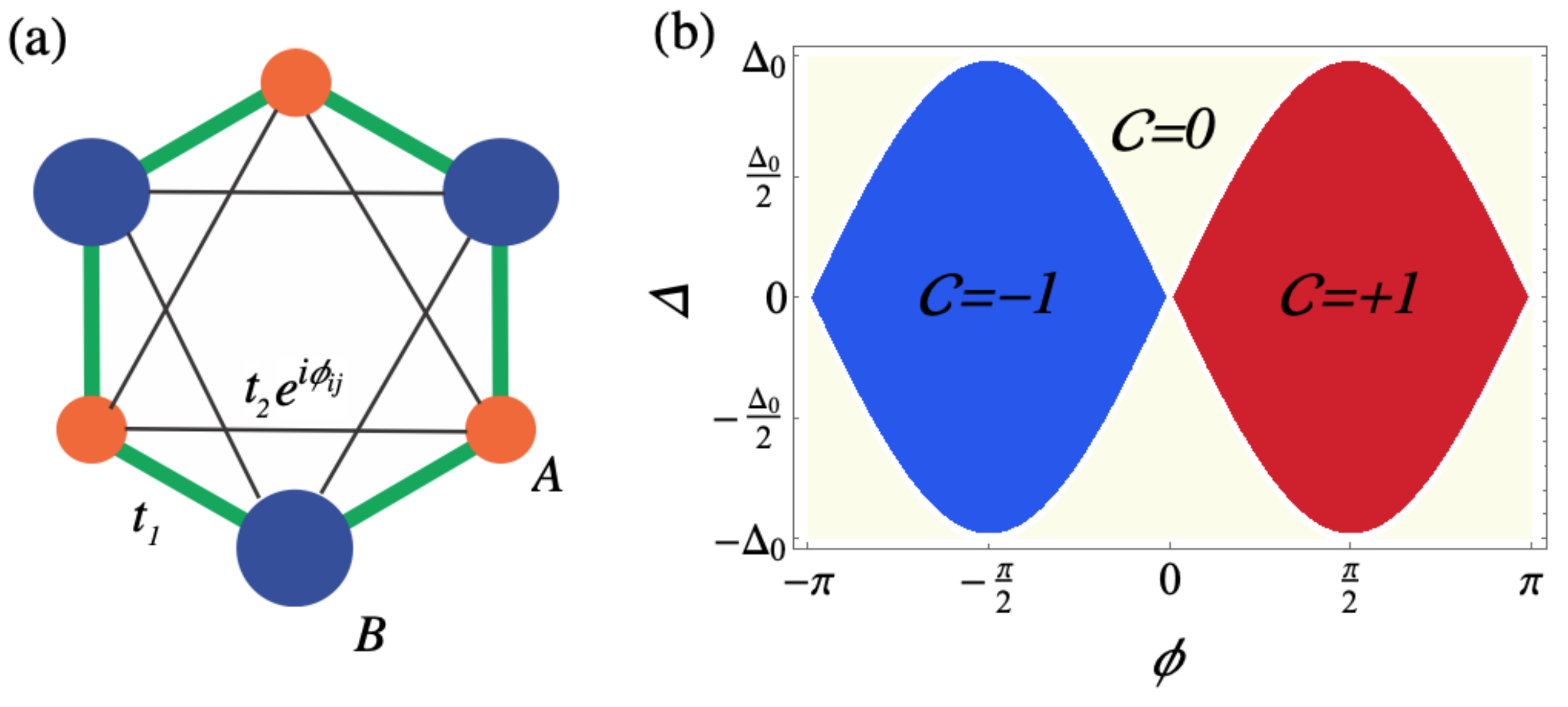}
\end{center}
\caption{(a) Both the impurity and majority particles live in a honeycomb lattice with  
 nearest neighbor hopping with strength $t_1$. The majority particles in addition experience next-nearest neighbour hopping with 
strength $t_2$ and phase $\phi_{ij}$, and an energy off-set between the triangular sublattices of $A$ and $B$ sites. 
(b) The phase diagram of the Haldane insulator for the majority atoms. 
Here $\Delta_0=3^{3/2}t_2$.} 
\label{LatticeFig}
\end{figure}
Thus, the Hamiltonian for the impurity corresponds to the usual nearest neighbour tight-binding model for graphene \cite{Neto2009}, whereas the 
 Hamiltonian for the majority particles corresponds to the Haldane model~\cite{Haldane1988}:  
\begin{gather}
\hat H_0=-t_1\sum_{\sigma=\uparrow,\downarrow}\sum_{\langle i,j\rangle}\hat c_{iA\sigma}^\dagger \hat c_{jB\sigma}-t_2\sum_{\llangle i,j\rrangle}e^{i\phi_{ij}}\hat c_{iA\uparrow}^\dagger \hat c_{jB\uparrow}+\text{h.c.}\nonumber\\
+\Delta \sum_{i}(\hat c_{iA\uparrow}^\dagger \hat c_{iA\uparrow}-\hat c_{iB\uparrow}^\dagger \hat c_{iB\uparrow})
=\sum_{\mathbf k\alpha\sigma}\varepsilon_{\sigma\alpha}(\mathbf k)\hat\gamma^\dagger_{\mathbf{k}\alpha\sigma}\hat \gamma_{\mathbf{k}\alpha\sigma}.
\label{Hamiltonian}
\end{gather}
Here, $\hat c_{is\,\sigma}^\dagger$ creates an impurity/majority particle for $\sigma=\downarrow,\uparrow$ on the  $s=A/B$-site in unit cell $i$,  
and $t_1$ is the nearest 
 neighbor hopping matrix element, which  is taken to be the same for both kinds of particles.   In the Haldane model~\cite{Haldane1988}, the matrix elements for the 
 next-nearest-neighbor hopping of the majority particles have strength $t_2$ and phase $\phi_{ij}$; we have $\phi_{ij}\!=\!\phi$, or $\phi_{ij}\!=\!-\phi$, depending on whether the next-nearest 
 hopping process is performed in a clockwise or anti-clockwise fashion, respectively. The staggered sub-lattice potential $\Delta$ splits the  energy of the A  and B sites. 
 The second line of Eq.~\eqref{Hamiltonian} displays the diagonalized Hamiltonian, and introduces the
 operator $\hat\gamma^\dagger_{\mathbf{k}\alpha\sigma}$, which creates a particle in the single-particle 
 eigenstate of the Haldane ($\sigma\!=\!\uparrow$) or graphene  ($\sigma\!=\!\downarrow$) Hamiltonian, in 
 band $\alpha\!=\!(1,2)$ and with quasi-momentum ${\mathbf k}$ within the first Brillouin zone (BZ); the  energy of this state is $\varepsilon_{\sigma\alpha}(\mathbf k)$. 
 In the following, we assume that the majority particles fill the lowest band completely  
in the absence of interactions so that they form a topological band insulator.
The well-known topological phase diagram of the Haldane model, characterized by the Chern number ${\mathcal C}$, is shown in Fig.~\ref{LatticeFig}(b).

The impurity interacts with the majority particles via the contact potential 
\begin{align}
\hat H_\text{int}&=g\sum_{i}\sum_{s=A,B}\hat c_{is\uparrow}^\dagger \hat c_{is\downarrow}^\dagger \hat c_{is\downarrow} \hat c_{is\uparrow}\label{Interaction}\\
&=\frac gN
\sum_{\substack{\mathbf{k k' q}\\\alpha\beta\alpha'\beta'}}W^{\alpha\beta}_{\alpha'\beta'}({\mathbf k},{\mathbf k}',{\mathbf q})
\hat\gamma^\dagger_{\mathbf{k+q}\alpha\uparrow}\hat\gamma^\dagger_{\mathbf{k'-q}\alpha'\downarrow}\hat \gamma_{\mathbf{k'}\beta'\downarrow} 
\hat \gamma_{\mathbf{k}\beta\uparrow},\notag
\end{align} 
where $N$ is the number of unit cells in the lattice, and 
$W^{\alpha\beta}_{\alpha'\beta'}({\mathbf k},{\mathbf k}',{\mathbf q})$ gives 
the strength of the scattering of a majority/impurity particle in band $\beta/\beta'$ with momentum ${\mathbf k}/{\mathbf k}'$ 
into a majority/impurity particle in band $\alpha/\alpha'$ with momentum ${\mathbf k}+{\mathbf q}/{\mathbf k}'-{\mathbf q}$. A detailed expression 
for $W^{\alpha\beta}_{\alpha'\beta'}({\mathbf k},{\mathbf k}',{\mathbf q})$ is given in Supplemental Material~\cite{SM}.

{\it The polaron.-}  The interaction $\hat H_\text{int}$ results in the creation of a quasiparticle called the Fermi polaron, which  consists of the impurity ``dressed" by 
 a cloud of fermionic majority atoms. Such polarons have been studied intensely 
in the absence of a lattice using cold atom systems~\cite{Schirotzek2009,Kohstall2012,Koschorreck2012,Scazza2017,Massignan2014}. 
 The polaron ground state wave function is well approximated by the so-called Chevy ansatz~\cite{Chevy2006} 
\begin{align}
\label{ChevyAnsatz}
|\psi_0\rangle=&(\sqrt{Z_0}+\sum_{\mathbf Q,\mathbf q,\alpha} M_{\mathbf Q,\mathbf q,\alpha}\hat \gamma^\dagger_{\mathbf{q}2\uparrow}\hat \gamma^\dagger_{\mathbf{Q}-\mathbf q\alpha\downarrow} \hat \gamma_{\mathbf{Q}1\uparrow}  \hat \gamma_{\mathbf{0}1\downarrow} )|\varphi_0\rangle\\ \nonumber
\equiv&\sqrt{Z_0}|\varphi_0\rangle+\sum_{\mathbf Q,\mathbf q,\alpha}M_{\mathbf Q,\mathbf q,\alpha}|\varphi_{\mathbf Q,\mathbf q,\alpha}\rangle,
\end{align}
where $Z_0$ is the  quasiparticle residue  
while  $|\varphi_0\rangle$ and $|\varphi_{\mathbf Q,\mathbf q,\alpha}\rangle$  are the 
non-interacting ground and excited states respectively.
The coefficients  $\sqrt{Z_0}$ and $M_{\mathbf Q,\mathbf q,\alpha}$ are obtained by minimising $\langle\hat{H}_0+\hat{H}_{\text{int}}-E\rangle$,
as explained 
in the Supplementary Material~\cite{SM}. The  
second term in Eq.~\eqref{ChevyAnsatz} describes the dressing of the impurity by particle-hole excitations of the majority particles from the valence to the 
conduction band while exciting the impurity to band $\alpha$. Since the Haldane bands 
 have non-trivial topological properties  for certain values of $(\phi,\Delta)$,  the polaron is a
  coherent mixture of a topologically-trivial impurity surrounded by a topological dressing cloud of majority particles. 
This raises the fascinating question of whether the polaron inherits some the topological properties of its dressing cloud and how one can characterize this phenomenon.  We stress that the
polarons stem from the genuine interacting nature of our composite system, and that such objects cannot be realized by simply connecting two non-interacting layers through hopping.

{\it External force and transverse current.-} 
We do not expect the topological properties of the polaron to be reflected in quantities such as its energy. 
 Inspired by the famous TKNN relation, linking the   transverse (Hall) conductivity to the Chern number~\cite{Thouless1982}, we instead examine the transverse response of the polaron to an 
external force.  The central question concerns the mechanism by which the quantized Hall response of the majority   induces a polaronic Hall effect through interactions.

In order to measure the response of the polaron  (and not that of the bare impurity only), we take the force ${\mathbf F}({\mathbf r})=-\boldsymbol{\nabla} V({\mathbf r})$ 
 to act on both the impurity and the majority particles, i.e.~on both components of the polaron. The perturbation corresponding to the force is then 
\begin{align}
\hat H'(t)=\int\! d^2rV({\mathbf r})\hat{\rho}({\mathbf r},t),
\label{Perturbation}
\end{align}
where $\hat{\rho}=\hat{\rho}_{\uparrow}+\hat{\rho}_{\downarrow}$ is the total density of the system.
For concreteness, we consider a uniform force $F_y$ in the $y$-direction, i.e. $\mathbf F(\mathbf r)\!=\!-F_y\mathbf e_y$. 
The transverse Hall conductivity of the polaron, $\sigma_{xy}^{P}$, then determines  the induced current density 
along the $x$-direction,  according to  $\langle\hat{j}_{x\downarrow}\rangle={\sigma_{xy}^{P}}\cdot (-F_y)$. 
The homogeneous current densities of the two components are  given by  the operators
\begin{align}
\hat{\mathcal{\mathbf  j}}_\sigma
=\frac1N\sum_{\mathbf k}\hat{\vec{\mathcal J}}(\mathbf k)
=\frac1N\sum_{\mathbf k}\sum_{\alpha\beta}\hat\gamma^\dagger_{\mathbf{k}\beta\sigma}\vec{\mathcal J}_{\sigma\beta\alpha}(\mathbf k)
\hat\gamma_{\mathbf{k}\alpha\sigma} ,
\label{CurrentOperator}
\end{align}  
 where the quantity $\vec{{\mathcal J}}_{\sigma\beta\alpha}(\mathbf k)$, giving the current operator in the eigenbasis of the graphene and 
Haldane Hamiltonians, is given in the Supplemental Material \cite{SM}.

The transverse current of the polaron due to the force Eq.~\eqref{Perturbation}  
can be written in terms of the current-current correlation function within linear response. 
 As shown in Ref.~\cite{SM}, we have 
\begin{align}
{\sigma}_{xy}^P=\lim_{\omega\rightarrow 0}-\frac{{\mathcal{ P}_{xy}}(\omega)}{i\omega} ,
\label{Identity}
\end{align}
where  ${\mathcal P}_{xy}(\omega)$ is the Fourier transform of the current-current correlation function  
 $\mathcal P_{xy}(t-t')= -iN\theta(t-t')\langle \psi_0|[\hat j_{x\downarrow}(t),\hat j_{y\uparrow}(t')+\hat j_{y\downarrow}(t')]|\psi_0\rangle$, with $\theta(t)$ the Heaviside function.  As we will show below, the transverse conductivity in Eq.~\eqref{Identity} 
  encodes the topological properties of the impurity dressed by the topological cloud in Eq.~\eqref{ChevyAnsatz}.  Besides, the longitudinal transport exhibits Bloch oscillations of the polaron~\cite{Grusdt2014}.

{\it Composite Berry curvature.-} 
The Berry curvature is an essential ingredient for understanding non-interacting Chern insulators~\cite{Hasan2010,Qi2011,Bernevig2013}. Likewise, 
the Lehmann representation can be used
to express the  transverse conductivity of the polaron in Eq.~\eqref{Identity} as 
an integral over a ``composite" Berry curvature,  
$\sigma_{xy}^P=\sum_{\mathbf Q}\mathcal{B}_{\downarrow\uparrow}(\mathbf Q)$, where~\cite{SM} 
\begin{align}
&\mathcal{B}_{\sigma\sigma'}(\mathbf Q)=-i\sum_{\mathbf k_1,\mathbf k_2,\mathbf q}\sum_{\alpha,\beta,\alpha',\beta',\alpha''}\label{CompositeBerry}\\
&\frac{\langle \psi_0|\hat{\mathcal J}^{x}_{\sigma\alpha\beta}(\mathbf k_1)|\psi_{\mathbf Q, \mathbf q,\alpha''}\rangle \langle \psi_{\mathbf Q, \mathbf q,\alpha''}|\hat{\mathcal J}^{y}_{\sigma'\alpha'\beta'}(\mathbf k_2)|\psi_0\rangle-x\leftrightarrow y}
{(E_0-E_{\mathbf Q, \mathbf q})^2}.\notag
\end{align} 
 Here $|\psi_{\mathbf Q, \mathbf q,\alpha}\rangle$ is an interacting excited state of the polaron with energy $E_{\mathbf Q, \mathbf q}$, which is adiabatically connected to the non-interacting excited state 
$|\varphi_{\mathbf Q, \mathbf q,\alpha}\rangle$ defined in Eq.~\eqref{ChevyAnsatz}.   Importantly, the quantity $\mathcal{B}_{\sigma\sigma'}(\mathbf Q)$, which describes the 
Berry curvature of an excitation involving spins $\sigma$ and $\sigma'$ with total momentum $\mathbf Q$, corresponds to the Berry curvature of the polaron in Eq.~\eqref{ChevyAnsatz} when setting $\sigma=\uparrow$ and $\sigma'=\downarrow$. 
This new quantity emerges as a consequence of the combination of the many-body nature of the polaron and the underlying topological band structure of the majority particles.
It  is easy to show that $\mathcal{B}_{\uparrow\uparrow}(\mathbf Q)$ on the other hand 
recovers the usual expression for the Berry curvature of the Haldane model~\cite{SM}.

 \begin{figure}
 \includegraphics[width=\columnwidth]{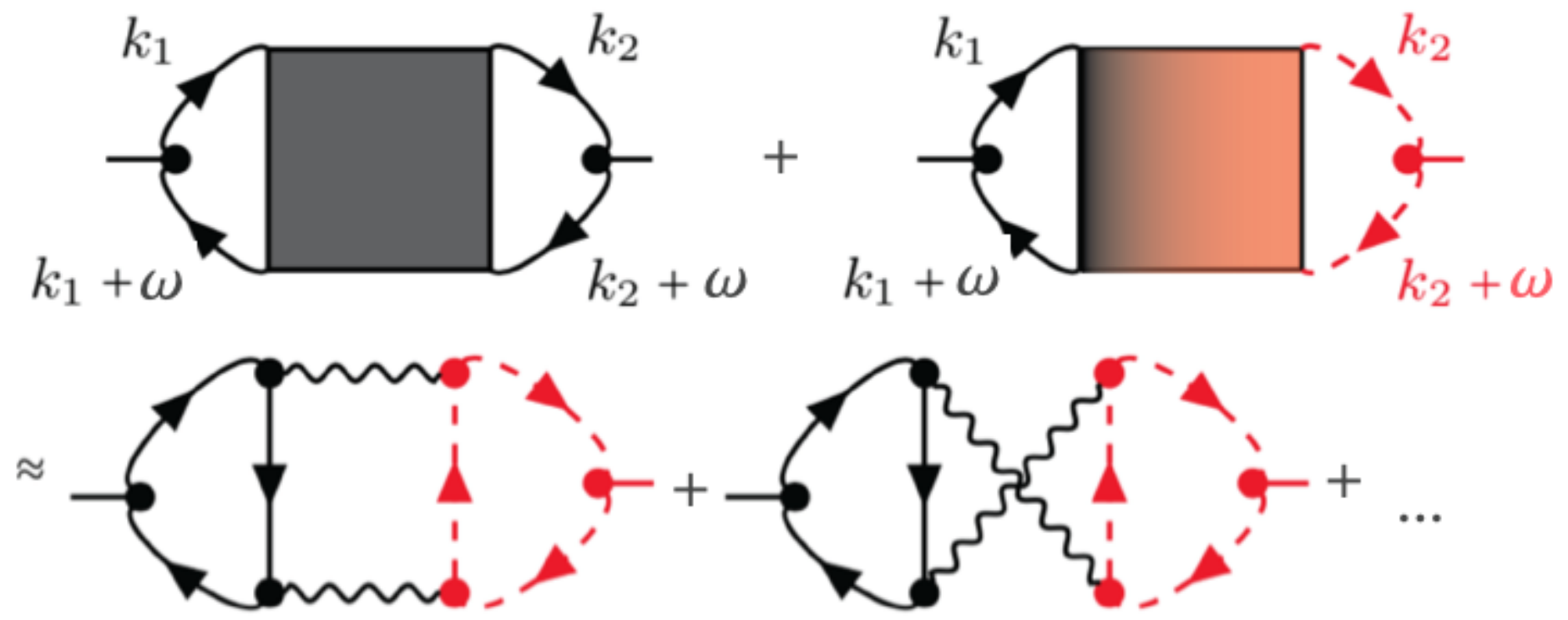}
   \caption{(Top) Diagrammatic representation of  the transverse conductivity of the polaron. (Bottom) Second-order non-zero diagrams.  Black solid/red dashed lines denote the 
  impurity/majority $\sigma=\downarrow,\uparrow$ Green's function and wavy lines the interaction. 
   }
 \label{FigDiagrams1}
 \end{figure}

{\it Diagrammatic analysis.-} We now use a diagrammatic analysis to calculate   the transverse conductivity Eq.~\eqref{Identity}. 
This allows us to include the interaction in a systematic way using perturbation theory in the coupling strength $g$.  The calculation is equivalent to 
using the many-body Chevy ansatz in Eq.~\eqref{ChevyAnsatz} to evaluate the composite Berry curvature in Eq.~\eqref{CompositeBerry}
 up to second-order in $g$. 
 The current-current correlation  function is illustrated diagrammatically in Fig.~\ref{FigDiagrams1}.
We assume zero temperature so that the polaron is initially in its ground state $|\psi_0\rangle$ with  zero momentum. 
Since the Berry curvature vanishes  for the energy bands of graphene, it follows that $\sigma_{xy}^P=0$ 
 when there are no interactions. The first order diagrams also give no contribution to the transverse conductivity, as they 
 correspond to a simple Hartree energy shift of the impurity energy. The first non-vanishing contribution to the transverse 
 conductivity is therefore second order in $g$, and it is given by the diagrams shown in  the lower panel of Fig.~\ref{FigDiagrams1}. 
They correspond to the  contribution 
 \begin{widetext}

\begin{gather}
{\mathcal P}_{xy}(\omega)=g^2\sum_{k_1k_2k_3}{\mathcal J}^x_{\downarrow\beta'\alpha'}({\mathbf k}_2){\mathcal G}_{\downarrow\beta'}(k_2+\omega)
{\mathcal G}_{\downarrow\alpha'}(k_2){\mathcal G}_{\downarrow\kappa'}(k_3+k_2)\left[
W^{\beta\kappa}_{\kappa'\beta'}({\mathbf k}_1+{\mathbf k}_3,{\mathbf k}_2,-{\mathbf k}_3)
W^{\kappa\alpha}_{\alpha'\kappa'}({\mathbf k}_1,{\mathbf k}_3+{\mathbf k}_2,{\mathbf k}_3)\times
\right.\nonumber\\
\left. 
 {\mathcal G}_{\uparrow\kappa}(k_1+k_3)
+W^{\beta\kappa}_{\alpha'\kappa'}({\mathbf k}_1-{\mathbf k}_3,{\mathbf k}_3+{\mathbf k}_2,{\mathbf k}_3)
W^{\kappa\alpha}_{\kappa'\beta'}({\mathbf k}_1,{\mathbf k}_2,-{\mathbf k}_3){\mathcal G}_{\uparrow\kappa}(k_1-k_3+\omega)\right]
{\mathcal G}_{\uparrow\alpha}(k_1){\mathcal G}_{\uparrow\beta}(k_1+\omega)
{\mathcal J}^y_{\uparrow\beta\alpha}({\mathbf k}_1),
\label{LongDiagramExpression}
\end{gather}
\end{widetext}
where $\sum_k\equiv T \sum_{\omega_n}\sum_{\mathbf k}$ is a shorthand notation for a summation over a Matsubara frequency and integration over a 2D momentum 
${\mathbf k}$ inside the BZ,  $k\equiv({\mathbf k},i\omega_n)$, and there is a summation over repeated band indices. 
The non-interacting Green's function for a  $\sigma$ particle in band $\alpha$ is 
${\mathcal G}_{\sigma\alpha}(k)^{-1}=i\omega_n-\varepsilon_{\sigma\alpha}(\mathbf k)$. 
 In Ref.~\cite{SM},  we provide all first- and second-order diagrams for the current-current  correlation function and  evaluate the three 
Matsubara sums in Eq.~\eqref{LongDiagramExpression} analytically. As usual, we add a positive infinitesimal part to the frequency $\omega$ 
 to get the retarded correlation function ${\mathcal P}_{xy}(\omega).$ 
\begin{figure*}
\includegraphics[width=7.0in,height=2.5in]{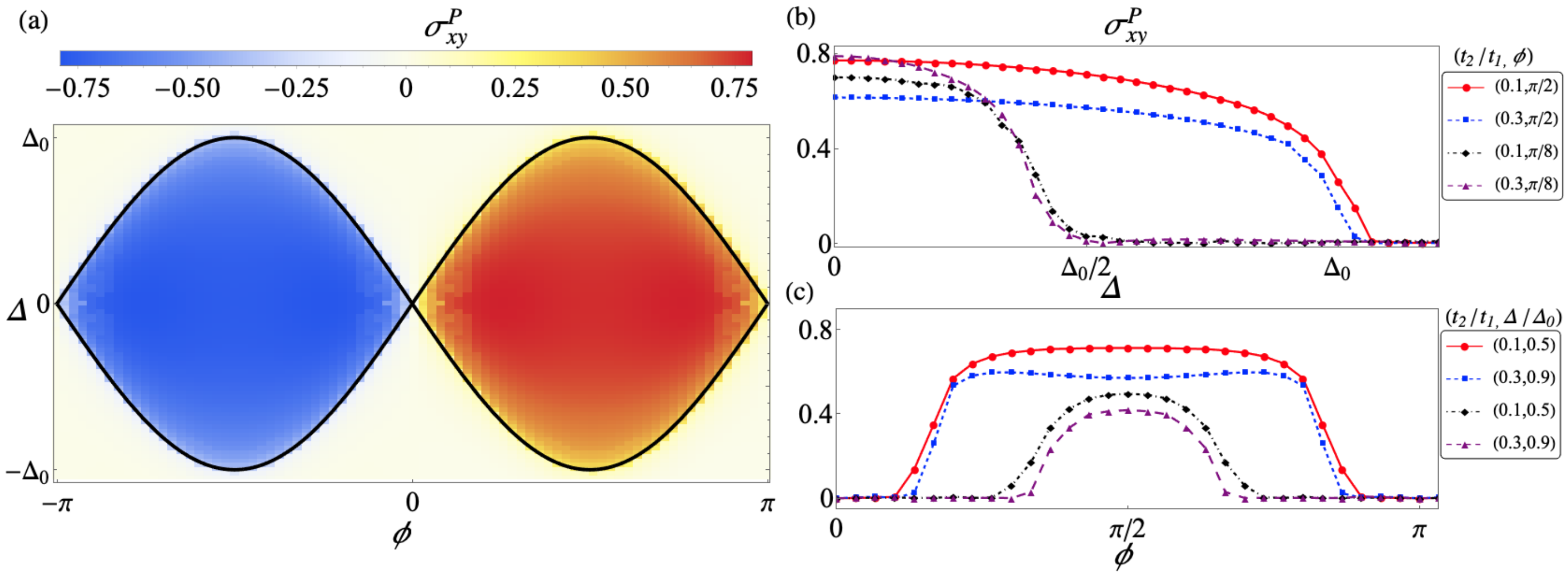}
\caption{(a) Transverse conductivity $\sigma_{xy}^P$ of the polaron in units of $g^2/Ng_0^2$ with $g_0=(2\pi a)^2/(3\sqrt{3}/2)t_1$, as a function of $\phi$ and $\Delta$ for $t_2=0.1t_1$. 
 The solid lines give the boundaries $\Delta=\pm 3^{3/2}t_2\sin\phi$ for the topological phase of the majority atoms. Panels (b) and (c) depict $\sigma_{xy}^P$ for fixed values of $(t_2/t_1,\phi)$ and $(t_2/t_1,\Delta/\Delta_0)$ respectively.   }
\label{Phasediagram0.1}
\end{figure*} 
  Note that our second order calculation is conserving~\cite{Baym1961}, which is a major challenge for 
arbitrary interaction strengths~\cite{Cotlet2018}.

The fact that the first non-zero contribution to the transverse conductivity is proportional to  $g^2$ 
can be understood as follows. The transverse current of the polaron is caused by the drag exerted by its dressing cloud. 
This drag is proportional to the scattering rate between the impurity and the majority particles, which again 
is proportional to the scattering cross section scaling as $g^2$. 

 \begin{figure}[h]
\begin{center}
\includegraphics[width=\columnwidth]{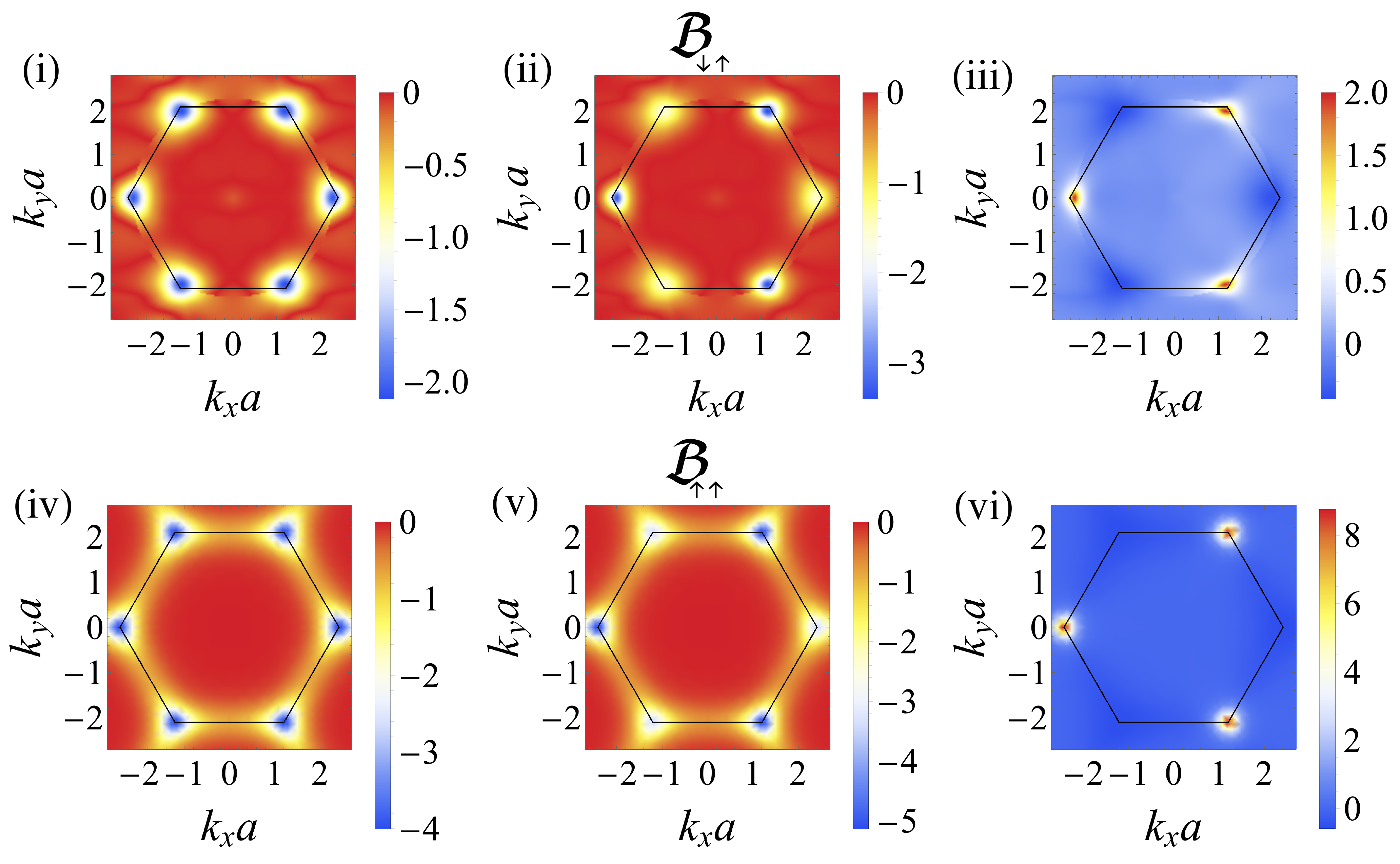}
\end{center}
\caption{Panels (i)-(iii) The composite Berry curvature $\mathcal{B}_{\downarrow\uparrow}(\mathbf Q)$ of the polaron for  (i) $(\phi,\Delta)=(-\pi/2,0)$, (ii) $(\phi,\Delta)=(-\pi/2,\Delta_0/3)$, and (iii) $(\phi,\Delta)=(-\pi/2,5\Delta_0/3)$.  
   Panels (iv)-(vi) below show  the corresponding Berry curvature $\mathcal{B}_{\uparrow\uparrow}(\mathbf Q)$ for the majority for the same values of $(\phi,\Delta)$.} 
\label{Fig4}
\end{figure}
 
{\it Results.-} We now discuss our main results shown in Figs.~\ref{Phasediagram0.1} and \ref{Fig4}. The  transverse conductivity of the ground state polaron $\sigma_{xy}^{P}$ 
 is obtained by evaluating the diagrammatic expression in Eq.~\eqref{LongDiagramExpression} numerically, assuming that  the impurity remains in the lower band.  
Figure~\ref{Phasediagram0.1}(a) shows that the transverse conductivity of the polaron has the same sign as that of the majority particles, as given by their Chern number ${\mathcal C}$. 
Moreover, $\sigma_{xy}^P$ vanishes  when the majority particles are in a trivial phase  (${\mathcal C}\!=\!0$).  
Thus, the polaron inherits the Hall-type transport properties of its dressing cloud, which is an effect solely due to interactions and which is deeply rooted in the topology of the underlying Chern insulator; this reflects the transverse drag that the majority particles impose on the impurity.

 Macroscopically, one could anticipate that $\sigma_{xy}^P\!=\!0$  whenever the majority particles are  in the trivial phase  (with no net transverse current). 
 However, this is not so obvious microscopically, as the individual majority particles   exhibit a 
 transverse motion also in the trivial phase due to the non-zero 
  Berry curvature of the  Haldane band~\cite{Xiao2010,Karplus1954}; since the impurity-majority scattering rate   depends on the quantum states involved, this could lead to a net transverse drag on the impurity.  We attribute the vanishing of $\sigma_{xy}^P$ to the fact that, in the trivial phase, one can use a single gauge to describe the majority particles, and hence, to the polaron eigenstates in Eq.~\eqref{CompositeBerry}. 
  
While the transverse conductivity of the polaron  is intimately related to the topological properties of its  dressing cloud, it is not quantised:  as shown in Fig.~\ref{Phasediagram0.1}(b)-(c), it varies  slightly as $\phi$ and $\Delta$ are changed, leading to a saddle-point-like surface. This reflects the composite many-body nature of the polaron,  whose transport and geometric properties arise as a combination of the topological properties associated with the majority and a series of non-universal features (e.g., interactions). 
To further investigate the geometric properties of the polaron, we present 
 its composite Berry curvature $\mathcal{B}_{\downarrow\uparrow}(\mathbf Q)$, and 
 the corresponding Berry curvature $\mathcal{B}_{\uparrow\uparrow}(\mathbf Q)$ of the  populated Haldane band  in Fig.~\ref{Fig4} (i)-(vi), for various values of $(\phi,\Delta)$.  Figures \ref{Fig4} (i)-(ii) and (iv)-(v) correspond to the  topological phase with ${\mathcal C}=-1$, and 
Figs.~\ref{Fig4} (iii) and (vi) correspond to the topologically trivial phase.  One finds that 
 the composite Berry curvature $\mathcal{B}_{\downarrow\uparrow}(\mathbf Q)$ closely mimics the Berry curvature 
$\mathcal{B}_{\uparrow\uparrow}(\mathbf Q)$ of the  Haldane band, and that it inherits all its
asymmetric features. This shows that the \emph{geometric} properties of the Haldane model are  faithfully
mapped onto the polaron, at the microscopic level of the polaron wave function. 
We point out that the precise shape of $\mathcal{B}_{\downarrow\uparrow}(\mathbf Q)$ differs from $\mathcal{B}_{\uparrow\uparrow}(\mathbf Q)$,  which indicates how the polaron Hall conductivity deviates from the quantized value experienced by the majority.

A non-zero temperature will reduce or smoothen the jump of the polaron's transverse conductivity at the topological transition, due to the thermal population of the excited band, but a well-defined feature should remain visible as long as the temperature is well below the band gap. To quantify more precisely this effect, our diagrammatic formalism may be extended by means of finite temperature Green's functions.  
 
{\it  Concluding remarks}.- The intricate interplay between many-body physics and topology discussed in this paper can be studied using present cold-atom technology. Polarons  have
been  systematically  investigated by several groups~\cite{Schirotzek2009,Koschorreck2012,Kohstall2012, Scazza2017,Jorgensen2016,Hu2016}.  Moreover, 
 the Haldane model has  been realized  and the Berry curvature of its energy bands   observed, using  optical lattices~\cite{Jotzu2014,Duca2015,Flaschner2016}. 
The Hall conductivity  of neutral atoms can be measured  through  transverse drift dynamics~\cite{Price2012,Dauphin2013,Deng2014,Aidelsburger2015,Price2016,anderson2017optical} 
 or via circular dichroism~\cite{Asteria2018}. We estimate the transverse velocity $v_{x\downarrow}$ of the impurity using $j_{x\downarrow}=n_{\downarrow}v_{x\downarrow}=\sigma_{xy}^PF_y$, where $n_{\downarrow}\sim1/Na^2$ is the impurity density   with $a$ the lattice constant. From this, the transverse displacement of the impurity after time $\tau$ is 
 $\delta x_\downarrow=\sigma_{xy}^PF_y \tau/n_{\downarrow}\approx0.25(n_{\uparrow}g/6t_1)^2a^2F_y \tau$, where we have used a 
 typical value $\sigma_{xy}^P\simeq0.8 g^2/Ng_0^2$, see  Fig.~\ref{Phasediagram0.1}.
 Hence, for typical experimental times $\tau\approx 20-50\pi/t_1$ and $F_y\approx0.2-0.3t_1/a$~\cite{Jotzu2014}, the transverse impurity displacement  is  significant,
  i.e.\ $\delta x_\downarrow\approx 1-5a$ even when the coupling  $g$ is small compared to the ``graphene band width" $6t_1$, so that our perturbative calculation is reliable. 
In typical polaron experiments, the concentration of impurities is typically  $\lesssim 20\%$. For such concentrations, the effects of polaron-polaron interactions are negligible due to the incompressibility of the Fermi gas~\cite{Scazza2017}.

We showed that the transverse conductivity of the polaron scales as 
$g^2$, when the force  acts on both the impurity and its dressing cloud. In fact, our result also holds when the force acts on the dressing cloud only. 
If instead the force  acts only on the impurity, we expect the transverse conductivity to scale as $g^4$. First, the longitudinal motion of the impurity due to the external force induces a longitudinal drag on the majority particles, which scales with the scattering cross section 
$\propto g^2$. The Berry curvature of the Haldane bands will then cause a transverse drift of the majority particles~\cite{Xiao2010,Karplus1954}, 
which causes a drag back on the impurity  scaling with $g^2$, giving a total $g^4$ scaling.   For strong coupling, we expect the transverse current to saturate 
when the impurity binds a single Haldane particle to form a dimer state.

Our results open up the exciting perspective of studying interacting topological systems using  
quantum impurities in atomic gases as a  highly controllable probe.

\vspace{.5cm}
ACG and GMB acknowledge the support of the Villum Foundation. 
Work in Brussels was supported by the Fonds De La Recherche Scientifique (FRS-FNRS) (Belgium) and the ERC Starting Grant TopoCold.
PM acknowledges the Spanish MINECO (FIS2017-84114-C2-1-P) and 
the ``Ram\'on y Cajal'' program. 

\bibliography{CGMB_Haldane}

\newpage
\title{Supplemental Material:  Dropping an impurity into a Chern insulator: a polaron view on topological matter}
\author{A.\ Camacho-Guardian$^1$, N.\ Goldman$^2$, P.\ Massignan$^3$, and G.\ M.\ Bruun$^1$}
\affiliation{$^1$Department of Physics and Astronomy, Aarhus University, Ny Munkegade, DK-8000 Aarhus C, Denmark}
\affiliation{$^2$Center for Nonlinear Phenomena and Complex Systems,
Universit\'e Libre de Bruxelles, CP 231, Campus Plaine, 1050 Brussels, Belgium}
\affiliation{$^3$Departament de F\'isica, Universitat Polit\`ecnica de Catalunya, Campus Nord B4-B5, E-08034 Barcelona, Spain}

\date{\today}

\maketitle
\begin{widetext}
\section{Interaction and current  operators in the eigenbasis}
\label{Interactionancurrentoperators}
The operator $\hat \gamma_{{\mathbf k}\alpha\sigma}^\dagger$ creates a spin $\sigma=\uparrow,\downarrow$ 
particle in an eigenstate of $\hat H_0$ with momentum ${\mathbf k}$ and band index $\alpha=1,2$. The unitary transformation between these 
operators and the momentum eigenstate operators for the $s= A /B$ sites, 
 $\hat c_{\mathbf ks\sigma}=N^{-1/2}\sum_j\hat c_{js\sigma}\exp(i{\mathbf k}\cdot{\mathbf r}_j)$, is
 \begin{align}\tag{S1}
\begin{bmatrix} \hat c_{\mathbf kA\sigma} \\ \hat c_{\mathbf k B\sigma} \end{bmatrix}=
{\mathcal U}_\sigma({\mathbf k})
\begin{bmatrix} \hat \gamma_{\mathbf k1\sigma} \\ \hat \gamma_{\mathbf k2\sigma} \end{bmatrix}.
\end{align}
This transformation for the contact interaction straightforwardly yields the second line in Eq.~(2) with 
\begin{equation}\tag{S2}
W^{\alpha\beta}_{\alpha'\beta'}({\mathbf k},{\mathbf k}',{\mathbf q})=\sum_{n=1}^2{\mathcal U}^*_{\uparrow n\alpha}({\mathbf{k+q}})
{\mathcal U}^*_{\downarrow n\alpha'}({\mathbf{k'-q}}){\mathcal U}_{\downarrow n\beta'}({\mathbf k'}){\mathcal U}_{\uparrow n\beta}({\mathbf k}).
\end{equation}

Using the same transformation, we obtain the  current operators for the impurity and majority particles in the eigenbasis as 
\begin{gather}\tag{S3}
\hat{\mathcal{\mathbf  j}}_\sigma
=\frac1N\sum_{\mathbf k}\begin{bmatrix} \hat c_{\mathbf kA\sigma}^\dagger & \hat c_{\mathbf kB\sigma}^\dagger  \end{bmatrix} 
\nabla_{\mathbf k}\mathcal{H}_\sigma({\mathbf k})\begin{bmatrix}  \hat c_{\mathbf kA\sigma} \\  \hat c_{\mathbf kA\sigma} \end{bmatrix}
=\frac1N\sum_{\mathbf k}
\begin{bmatrix} \hat \gamma_{\mathbf k1\sigma}^\dagger & \hat \gamma_{\mathbf k2\sigma}^\dagger  \end{bmatrix} 
{\mathcal U}_\sigma^\dagger({\mathbf k})\nabla_{\mathbf k}\mathcal{H}_\sigma({\mathbf k}){\mathcal U}_\sigma({\mathbf k})
\begin{bmatrix}  \hat \gamma_{\mathbf k1\sigma} \\  \hat \gamma_{\mathbf k2\sigma} \end{bmatrix},
\end{gather} 
which defines the quantity $\vec{{\mathcal J}}_{\sigma\beta\alpha}(\mathbf k)$ introduced in Eq.~(5).

\section{Linear Response}
\label{AppendixLehmann}
We now show that the transverse conductivity determined by the current-density correlations  can be written in terms of the usual current-current correlations. By means of the continuity equation
\begin{equation}\tag{S4}
\frac{d\hat{\rho}(\mathbf r,t)}{dt}+\nabla\cdot \hat{\mathbf j}(\mathbf r,t)=0,
\end{equation}
the Hamiltonian for the perturbation in terms of the current operator is
\begin{gather}\tag{S5}
\label{h1}
\hat{H}'(t)=\int\! d\mathbf rV({\mathbf r})\hat{\rho}({\mathbf r},t)=\int d\mathbf r \int_{-\infty}^tdt'\frac{d\hat{\rho}(\mathbf r,t')}{dt'}V(\mathbf r)
=-\int d\mathbf r\int_{-\infty}^tdt'  \mathbf F(\mathbf r)\cdot \hat{\mathbf j}(t'),
\end{gather}
where $\hat{\rho}=\hat{\rho}_\uparrow+\hat{\rho}_\downarrow$ and $\hat{\mathbf j}=\hat{\mathbf j}_\uparrow+\hat{\mathbf j}_\downarrow$ denote the total density and current operators.
The transverse conductivity containing the linear terms in the perturbation $\hat{H}'(t)$ is then given by
\begin{gather}\tag{S6}
\sigma_{xy}^{P}(t-t')= -iN\theta(t-t')\langle [\hat j_{x\downarrow}(t),\hat{Q}_y(t')]\rangle,
\label{WeirdCorrelationFn}
\end{gather}
where $\hat{Q}_y(t')=\int_{-\infty}^{t'}\!dt''[\hat j_{y\downarrow}(t'')+\hat j_{y\uparrow}(t'')]$ and  $\langle\ldots\rangle$ denotes the thermal average. After some algebra and in the spectral representation the transverse conductivity is equal to
\begin{equation}\tag{S7}
\sigma_{xy}^{P}(t-t')=-iN\theta(t-t')\sum_n\left(\frac{e^{i(E_0-E_n)(t-t')}}{E_0-E_n}\langle \psi_0|\hat j_{x\downarrow}|\psi_n\rangle \langle \psi_n|\hat j_{y}|\psi_0\rangle+\frac{e^{-i(E_0-E_n)(t-t')}}{E_0-E_n}\langle \psi_0|\hat j_{y}|\psi_n\rangle \langle \psi_n|\hat j_{x\downarrow}|\psi_0\rangle \right),
\end{equation}
where $\{|\psi_n\rangle\}$ denotes the set of complete eigenvectors of the many-body Hamiltonian, with $E_0$ and $E_n$ the eigenenergies for the ground state and excited states respectively. Taking the Fourier transform and the static limit, the transverse conductivity $\sigma_{xy}^{P}=\sigma_{xy}^{P}(\omega=0)$ reads as
\begin{gather}\tag{S8}
\label{S8}
\sigma_{xy}^{P}=-i\sum_n\left(\frac{\langle \psi_0|\hat{j}_{x\downarrow}|\psi_n\rangle \langle \psi_n|\hat j_{y}|\psi_0\rangle-\langle \psi_0|\hat j_{y}|\psi_n\rangle \langle \psi_n|\hat{j}_{x\downarrow}|\psi_0\rangle}{(E_0-E_n)^2} \right).
\end{gather}
Let us notice here that $\sigma_{xy}^{P}=\lim_{\omega\rightarrow 0}-\frac{\mathcal P_{xy}(\omega)}{i\omega}$. That is, $\sigma_{xy}^P$ can be written in terms of the current-current correlations instead of the {\it unusual}  current-density correlation. This expression makes the link between neutral atoms in presence of a constant force and the usual formula for the transverse conductivity of particles with  charge $q$ subjected to a field $E_y=-\partial_tA_y$. Note that, using $qA_y=qE_y/i\omega=-F_y/i\omega$, we can also identify 
the current-density correlation function with the usual result in terms of the current-current correlation function, that is $\langle\hat{j}_{x\downarrow}\rangle=\lim_{\omega\rightarrow 0}{\mathcal P}_{xy}(\omega)\cdot (-qA_y(\omega))$.

\section{Diagrammatic analysis}
\label{AppendixFeynman}
Here we briefly discuss the diagrammatic approach for the current-current correlations. 
As schematised in Fig.~\ref{FigDiagrams}, the first term in the current-current correlator arises as the contribution of the force acting on the impurity, while the second term concerns the force acting on the majority. In the following we evaluate the corresponding Feynman diagrams using perturbation theory.

 \begin{figure}[h]
 \includegraphics[width=3.9in,height=1.025in]{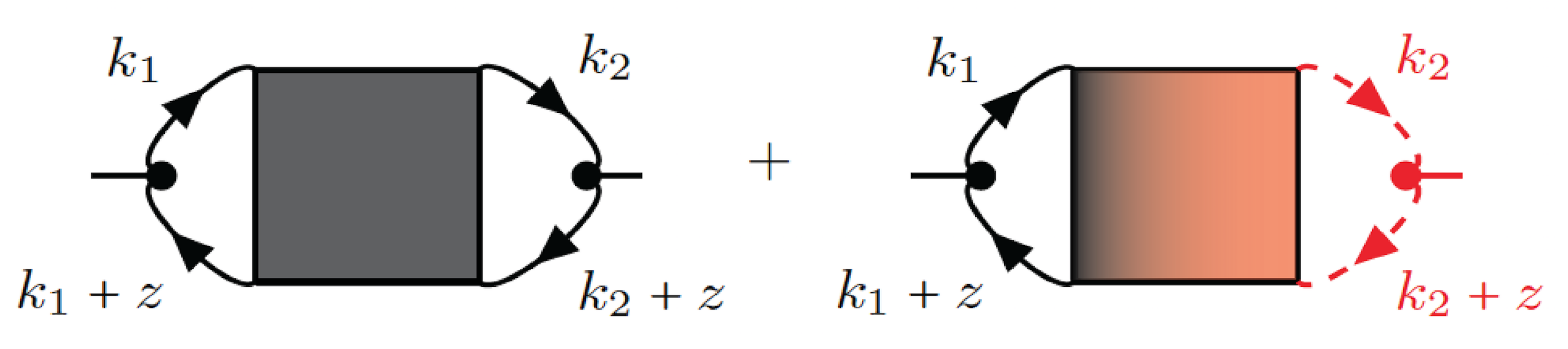}
     \caption{Diagrammatic representation of  the transverse conductivity of the polaron.  Black solid/red dashed lines denote the 
  impurity/majority $\sigma=\downarrow,\uparrow$ Green's function.
   }
 \label{FigDiagrams}
 \end{figure}

\centerline{{\bf Zero Order}}
The zero-order contribution schematised in Fig.~\ref{FigDiag0} is related to the Berry curvature of the honeycomb lattice. Since the Berry curvature
vanishes for the energy bands of the impurity, then the zero-order term vanishes exactly.

\begin{gather}\nonumber
i\mathcal{P}^{(0)}_{xy}(z)=\sum_{k}\text{Tr}\left[\mathcal J^x_{\downarrow}(\mathbf k)G_{\downarrow}(k)\mathcal J^y_{\downarrow}(\mathbf k) G_{\downarrow}(k+z)\right]
=\left[\frac{\mathcal J^x_{\downarrow12}(\mathbf k) J^y_{\downarrow21}(\mathbf k)}{z-(\varepsilon_{\downarrow,2}(\mathbf k)-\varepsilon_{\downarrow,1}(\mathbf k))}-\frac{\mathcal J^y_{\downarrow21}(\mathbf k) J^x_{\downarrow12}(\mathbf k)}{z+\varepsilon_{\downarrow,2}(\mathbf k)-\varepsilon_{\downarrow,1}(\mathbf k)}\right]=\mathcal{B}_{\downarrow\downarrow}({\mathbf k}),
\end{gather}

 \begin{figure}[h]
 \includegraphics[width=1.9in,height=1.0in]{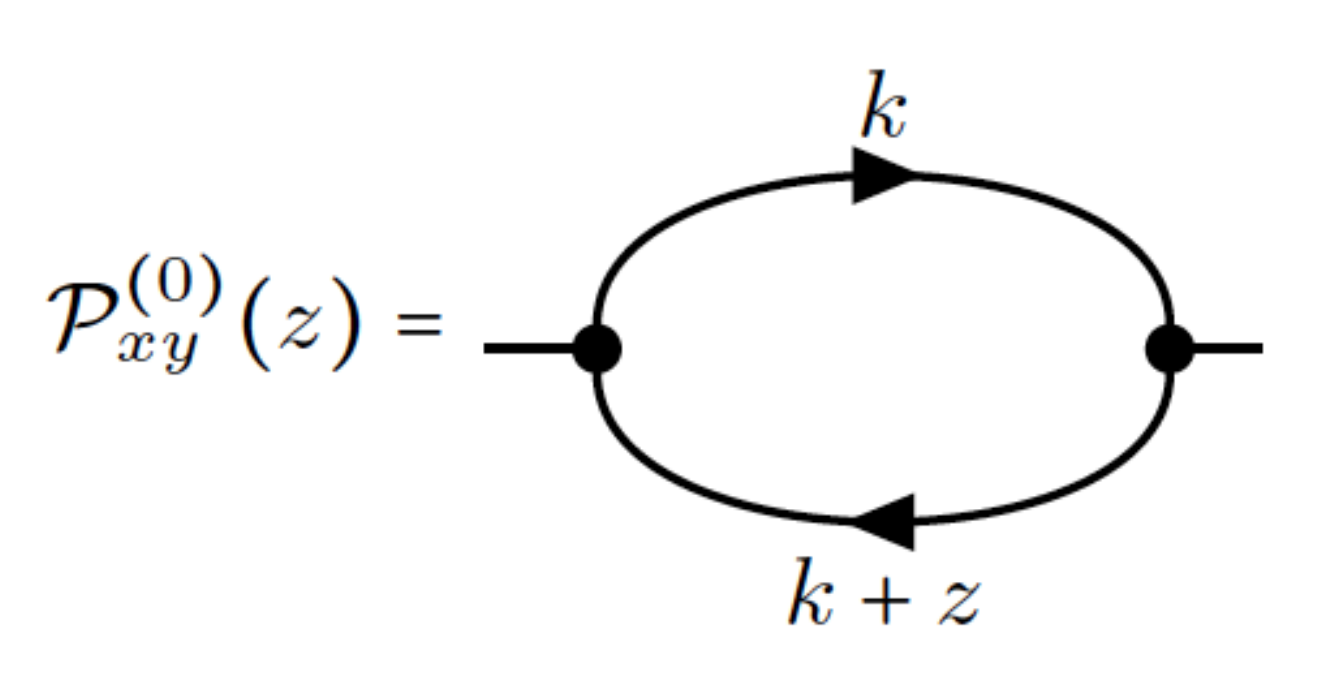}
   \caption{Zero-order contribution to the current-current correlation. This term vanishes for a single impurity.}
\label{FigDiag0}
 \end{figure}

\centerline{{\bf First order}}
  
  In Fig.~\ref{FigDiag1} we illustrate the only first-order diagram which is not simply a Hartree energy shift of the impurity energy. The explicit expression for such a diagram is
     \begin{gather}\tag{S9}
  \mathcal{P}_{xy}^{(1)}(z)=\sum_{ k, q}\left[ \mathcal J^x_{\downarrow\alpha'\beta'}(\mathbf k)G_{\downarrow\beta'}(k)G_{\downarrow\alpha'}(k+z) W^{\beta'\alpha'}_{\alpha,\beta}(\mathbf q,\mathbf k,0)\mathcal J^y_{\downarrow\alpha\beta}(\mathbf q)G_{\uparrow\beta}(q+z)G_{\uparrow\alpha}(q)\right].
   \end{gather}
 Since the transferred energy-momentum is always zero, the two sums decouple and yield
 a strictly zero contribution.
  \begin{figure}[h]
 \includegraphics[width=2.7in,height=.90in]{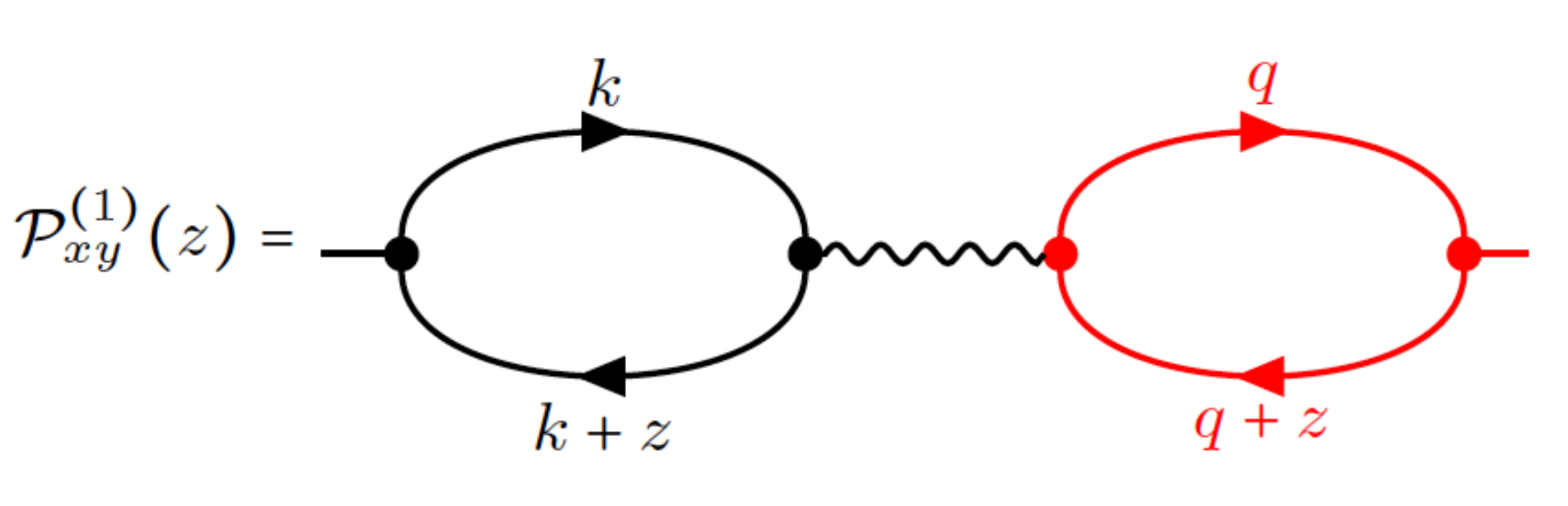}
   \caption{First order diagrams of the current-current correlation.}
 \label{FigDiag1}
 \end{figure}

\centerline{{\bf Second-order diagrams}}
The second-order diagrams which give a non-zero contribution to the transverse conductivity are shown in  Fig.~\ref{FigDiag2}.
We recall that the current-current correlations and the transverse conductivity are connected by the static limit $\sigma_{xy}^{P}=\lim_{\omega\rightarrow 0}\frac{i\mathcal{P}_{xy}(\omega)}{\omega}$. Therefore, we focus on the terms that yield a non-zero contribution when the static limit is taken. After performing the Matsubara sums, the final expression for the non-vanishing terms in the static limit where the impurity remains in the lower band is given by
\begin{gather}\nonumber \lim_{\omega\rightarrow 0}\frac{i\mathcal{P}_{xy}^{(2)}(\omega)}{\omega}=\lim_{\omega\rightarrow 0}\frac{1}{\omega}\sum_{\mathbf k,\mathbf q}\left[\frac{\mathcal J^x_{\downarrow11}(\mathbf q)\mathcal J^x_{\uparrow12}(\mathbf k)W^{21}_{11}(\mathbf k,\mathbf q,\mathbf 0)W^{11}_{11}(\mathbf k+\mathbf q,\mathbf 0,-\mathbf q)}{(\varepsilon_{1,\uparrow}(\mathbf k+\mathbf q)-\varepsilon_{2,\uparrow}(\mathbf k)-\varepsilon_{1,\downarrow}(\mathbf q)+\omega)(\varepsilon_{2,\uparrow}(\mathbf k)-\varepsilon_{1,\uparrow}(\mathbf k)-\omega)(\varepsilon_{1,\uparrow}(\mathbf k+\mathbf q)-\varepsilon_{2,\uparrow}(\mathbf k)-\varepsilon_{1,\downarrow}(\mathbf q))}\right.\\ \nonumber
+\frac{\mathcal J^x_{\downarrow11}(\mathbf q)\mathcal J^x_{\uparrow21}(\mathbf k)W^{11}_{11}(\mathbf k,\mathbf q,\mathbf 0)W^{12}_{11}(\mathbf k+\mathbf q,\mathbf 0,-\mathbf q)}{(\varepsilon_{1,\uparrow}(\mathbf k+\mathbf q)-\varepsilon_{2,\uparrow}(\mathbf k)-\varepsilon_{1,\downarrow}(\mathbf q)-\omega)(\varepsilon_{2,\uparrow}(\mathbf k)-\varepsilon_{1,\uparrow}(\mathbf k)+\omega)(\varepsilon_{1,\uparrow}(\mathbf k+\mathbf q)-\varepsilon_{2,\uparrow}(\mathbf k)-\varepsilon_{1,\downarrow}(\mathbf q))}\\ \nonumber
+\frac{\mathcal J^x_{\downarrow 11}(\mathbf k)\mathcal J^x_{\uparrow12}(\mathbf q)W^{22}_{11}(\mathbf q,\mathbf 0,-\mathbf k)W^{21}_{11}(\mathbf q-\mathbf k,\mathbf k,\mathbf k)}{(\varepsilon_{1,\uparrow}(\mathbf q)-\varepsilon_{2,\uparrow}(\mathbf q-\mathbf k)-\varepsilon_{1,\downarrow}(\mathbf k))(\varepsilon_{1,\uparrow}(\mathbf q)-\varepsilon_{2,\uparrow}(\mathbf q-\mathbf k)-\varepsilon_{1,\downarrow}(\mathbf k)+\omega)(\varepsilon_{1,\uparrow}(\mathbf q)-\varepsilon_{2,\uparrow}(\mathbf q)+\omega)} \\ \nonumber
\left.+\frac{\mathcal J^x_{\downarrow 11}(\mathbf k)\mathcal J^x_{\uparrow12}(\mathbf q)W^{12}_{11}(\mathbf q,\mathbf 0,-\mathbf k)W^{22}_{11}(\mathbf q-\mathbf k,\mathbf k,\mathbf k)}{(\varepsilon_{1,\uparrow}(\mathbf q)-\varepsilon_{2,\uparrow}(\mathbf q-\mathbf k)-\varepsilon_{1,\downarrow}(\mathbf k))(\varepsilon_{1,\uparrow}(\mathbf q)-\varepsilon_{2,\uparrow}(\mathbf q-\mathbf k)-\varepsilon_{1,\downarrow}(\mathbf k)-\omega)(\varepsilon_{1,\uparrow}(\mathbf q)-\varepsilon_{2,\uparrow}(\mathbf q)-\omega)}\right].\\ \tag{S10}
\label{S10}
\end{gather}
  \begin{figure}[h]
 \includegraphics[width=4.2in,height=1.20in]{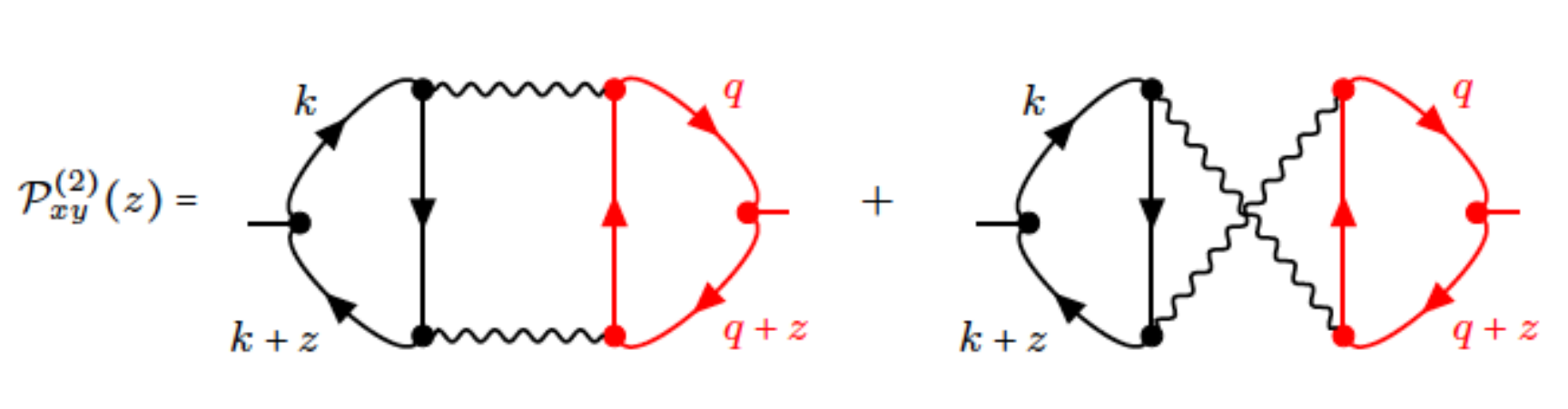}
   \caption{Leading diagrams of the current-current correlations}
\label{FigDiag2}
 \end{figure}
For simplicity, in the notation in Eq.~\eqref{S10} the energy of the impurity is shifted by $\varepsilon_{1,\uparrow}(\mathbf 0)$. 

\centerline{{\bf Chevy ansatz and second-order perturbation theory}}
For the Fermi polaron, a variational minimization of the energy over the Chevy ansatz is equivalent to the ``non self-consistent T-matrix approximation'' (NSCT) diagrammatic scheme, which takes into account all Feynman diagrams with one particle-hole excitation, see Refs.~[45] and [46]. 

For the problem under study here, the minimization of the functional $\langle \hat H_0+\hat H_{\text{int}}-E\rangle$ with respect to the variational parameters $(\sqrt{Z_0})^*$ and  $(M _ { \mathbf { Q } , \mathbf { q } , \alpha } ^ { 0 })^* $ of the Chevy ansatz yields the following coupled equations:
\begin{equation}
\label{S11}
\tag{S11}
\begin{split}
M _ { \mathbf { Q } , \mathbf { q } , \alpha } ^ { 0 }= & \frac{\frac{g}{N}\chi_{\alpha}(\mathbf Q,\mathbf q)}{E-(\varepsilon_{2\uparrow}(\mathbf q)+\varepsilon_{\alpha\downarrow}(\mathbf Q-\mathbf q)-\varepsilon_{1\uparrow}(\mathbf Q))},\\  
E\sqrt{Z_0}= & \frac{g}{N}\sum_{\mathbf q}\sqrt{Z_0}W_{11}^{11}(\mathbf q,\mathbf 0,\mathbf 0)+\frac{g}{N}\sum_{\mathbf q,\mathbf Q}W^{
12}_{\alpha\,1}(\mathbf q,\mathbf Q-\mathbf q,\mathbf Q-\mathbf q)M^0_{\mathbf Q,\mathbf q,\alpha},
\end{split}
\end{equation}
where we have introduced the auxiliary function
\begin{gather}
\chi_{\alpha}(\mathbf Q,\mathbf q)=W^{21}_{\alpha\,1}(\mathbf Q,\mathbf 0,\mathbf q-\mathbf Q)\sqrt{Z_0}+\sum_{\mathbf p,\alpha'}W^{22}_{\alpha'\alpha}(\mathbf p,\mathbf Q-\mathbf p,\mathbf q-\mathbf p)M _ { \mathbf { Q } , \mathbf { p } , \alpha' } ^ { 0 }
+W^{11}_{\alpha'\alpha}(\mathbf Q,\mathbf p-\mathbf q,\mathbf p-\mathbf Q)M _ { \mathbf { p } , \mathbf { q } , \alpha '} ^ { 0 }.
\label{S12}
\tag{S12}
\end{gather}
These coupled equations may be solved iteratively in the coupling strength $g$.
In absence of interactions, the ground state is simply given by $\sqrt{Z_0}=1$ and $M _ { \mathbf { Q } , \mathbf { q } , \alpha } ^ { 0 }=0 $. 
Iterating to second order in $g$, the evaluation of the current-current correlation using the variational approach coincides with the current-current correlation using both,  the diagrammatic approach and the perturbed state of the form
\begin{multline}
\label{S13}
\tag{S13}
| \psi _ { 0 } \rangle = \left( 1 - \sum _ { \mathbf { Q } , \mathbf { q } , \alpha } \frac { 1 } { 2 } \frac { \left\langle \varphi _ { 0 } \left| \hat { H } _ { \mathrm { int } } \right| \varphi _ { \mathbf { Q } , \mathbf { q } , \alpha } \right\rangle \left\langle \varphi _ { \mathbf { Q } , \mathbf { q } , \alpha } \left| \hat { H } _ { \mathrm { int } } \right| \varphi _ { 0 } \right\rangle } { \left( E _ { 0 } - E _ { \mathbf { Q } , \mathbf { q } } \right) ^ { 2 } } \right) | \varphi _ { 0 } \rangle\\ \nonumber
+ \sum _ { \mathbf { Q } , \mathbf { q } , \alpha } \left( \frac { \left\langle \varphi _ { \mathbf { Q } , \mathbf { q } , \alpha } \left| \hat { H } _ { \mathrm { int } } \right| \varphi _ { 0 } \right\rangle } { E _ { 0 } - E _ { \mathbf { Q } , \mathbf { q } , \alpha } }+\sum _ { \mathbf { Q } ^ { \prime } , \mathbf { q } ^ { \prime } , \alpha ^ { \prime } } \frac { \left\langle \varphi _ { \mathbf { Q } , \mathbf { q } , \alpha } \left| \hat { H } _ { \mathrm { int } } \right| \varphi _ { \mathbf { Q } ^ { \prime } , \mathbf { q } ^ { \prime } , \alpha ^ { \prime } } \right\rangle \left\langle \varphi _ { \mathbf { Q } ^ { \prime } , \mathbf { q } ^ { \prime }  , \alpha ^ { \prime }} \left| \hat { H } _ { \mathrm { int } } \right| \varphi _ { 0 } \right\rangle } { \left( E _ { 0 } - E _ { \mathbf { Q } , \mathbf { q } , \alpha } \right) \left( E _ { \mathbf { Q } , \mathbf { q } , \alpha } - E _ { \mathbf { Q } ^ { \prime } , \mathbf { q } ^ { \prime } , \alpha ^ { \prime } } \right) }- 
\frac { 
\left\langle \varphi _ { \mathbf { Q } , \mathbf { q } , \alpha } \left| \hat { H } _ {\text{int} } \right| \varphi _ { 0 } \right\rangle 
\left\langle \varphi _ { 0 } \left| \hat { H } _ { \text {int} } \right| \varphi _ { 0 } \right\rangle
} 
{ \left( E _ { 0 } - E _ { \mathbf { Q } , \mathbf { q } , \alpha } \right) ^ { 2 } } \right) | \varphi _ { \mathbf { Q } ,\mathbf q , \alpha } \rangle \\ \nonumber
= \sqrt { Z _ { 0 } } | \varphi _ { 0 } \rangle + \sum _ { \mathbf { Q } , \mathbf { q } , \alpha } M _ { \mathbf { Q } , \mathbf { q } , \alpha } ^ { 0 } | \varphi _ { \mathbf { Q } , \mathbf { q } , \alpha } \rangle.
\end{multline}
This procedure agrees with second-order perturbation theory with one particle-hole excitation. The correspondence can be understood as a reminiscent of the equivalence between the Chevy ansatz and the diagrammatic NSCT.

 We remark that the Lehmann representation in Eq.~\eqref{S8} requires also the determination of the many-body excited states $ |\psi_{\mathbf Q,\mathbf q,\alpha}\rangle.$ The equation for the variational ansatz for these states have a similar structure than the expression for the ground state in Eq.~\eqref{S11}.  The variational ansatz for the excited states is 
 \begin{equation} 
  |\psi_{\mathbf Q,\mathbf q,\alpha}\rangle= |\varphi_{\mathbf Q,\mathbf q,\alpha}\rangle+M^{\mathbf Q,\mathbf q}_0 |\varphi_0\rangle+\sum_{\alpha'}\sum_{\mathbf Q',\mathbf q'} M^{\mathbf Q,\mathbf q}_{\mathbf Q',\mathbf q'}|\varphi_{\mathbf Q',\mathbf q',\alpha'}\rangle, 
 \tag{S14}
\end{equation}
where again we consider only one particle-hole excitations, with $M^{\mathbf Q,\mathbf q}_0$ and $M^{\mathbf Q,\mathbf q}_{\mathbf Q',\mathbf q'}$ the variational parameters. 

We conclude by noting that the diagrammatic scheme  provides a guide to extend the current-current evaluation for non-perturbative approaches within the indispensable conserving approximations (see Refs.~[49] and [51] in the main text).

\end{widetext}

 \end{document}